%% file: main.tex
\documentclass{article}

\usepackage[final]{include/neurips_2019}

\usepackage[utf8]{inputenc} %
\usepackage[T1]{fontenc}    %
\usepackage{hyperref}       %
\usepackage{url}            %
\usepackage{booktabs}       %
\usepackage{amsfonts}       %
\usepackage{nicefrac}       %
\usepackage{microtype}      %
\usepackage{hyperref}

\usepackage{comment}
\usepackage{amsmath}
\usepackage{amssymb}
\usepackage{amsthm}
\usepackage{siunitx}
\usepackage{wrapfig}
\usepackage{todonotes}
\usepackage{enumitem}
\usepackage{siunitx}
\usepackage{dsfont}
\usepackage{xcolor}
\usepackage{tablefootnote}
\usepackage{multirow}
\usepackage{changepage}
\usepackage{bbm,bm}
\usepackage{tabularx} 
\usepackage{pgfplots}
\usepgfplotslibrary{units}
\usepackage{tikz}
\usetikzlibrary{trees}
\usepackage{calc}
\usepackage{tikz-qtree}
\usepackage{wrapfig}
\usepackage{upgreek}
\newsavebox\CBox
\newcommand\hcancel[2][1.5pt]{%
  \ifmmode\sbox\CBox{$#2$}\else\sbox\CBox{#2}\fi%
  \makebox[0pt][l]{\usebox\CBox}%
  \rule[0.5\ht\CBox-#1/2]{\wd\CBox}{#1}}

\usepackage{verbatim}
\usepackage{caption}
\usepackage[bottom]{footmisc}

\newcommand{\Set}[1]{\mathcal{#1}}

\DeclareMathOperator*{\setintersect}{ \cap }
\DeclareMathOperator*{\Setintersect}{ \bigcap }

\DeclareMathOperator*{\Setand}{ \bigwedge }
\newcommand{\closer}[3]{{\kern-#1ex{#2}\kern-#3ex}}
\newtheorem{lemma}{Lemma}

\usepackage{algorithm}
\usepackage{algorithmicx}
\usepackage[noend]{algpseudocode}

\input{include/murphy}

\input{include/custom}

\title{Multi-Agent Common Knowledge\\Reinforcement Learning}

\author{
	Christian A. Schroeder de Witt\thanks{Equal contribution. Correspondence to Christian Schroeder de Witt <cs@robots.ox.ac.uk>}\ \ \thanks{University of Oxford, UK}
\qquad
  Jakob N. Foerster\normalsize{\textsuperscript{$*$}}\textsuperscript{\textdagger}\\\quad\\
  {\bf Gregory Farquhar\textsuperscript{\textdagger}\qquad Philip H. S. Torr\textsuperscript{\textdagger}\qquad Wendelin B\"ohmer\textsuperscript{\textdagger}\qquad Shimon Whiteson\textsuperscript{\textdagger}}
  \date{}
}

\begin{document}

\maketitle

\input{core/0_abstract_v2}

\input{core/1_introduction}

\input{core/a_luck}
\input{core/3_background}

\input{core/4_method_v2}

\input{core/6_experiments}

\input{core/2_related_work}

\input{core/8_conclusion}

\newpage
\input{core/ack}

\bibliographystyle{include/icml2019}
\bibliography{references}

\newpage
\appendix
\part*{Appendix}
\input{core/9_supplementary}
\input{core/common_knowledge}

\end{document}

%% file: include/murphy.tex
\newcommand{\myvec}[1]{\mathbf{#1}}

\newcommand{\vu}{\myvec{u}}

\newcommand{\vI}{\myvec{I}}

\newcommand{\union}{\cup}

\newcommand{\be}{\begin{equation}}
\newcommand{\ee}{\end{equation}}
\newcommand{\bea}{\begin{eqnarray}}
\newcommand{\eea}{\end{eqnarray}}
\newcommand{\beaa}{\begin{eqnarray*}}
\newcommand{\eeaa}{\end{eqnarray*}}

\DeclareMathAlphabet{\mathpzc}{OT1}{pzc}{m}{n}

%% file: include/custom.tex
\newcommand{\method}{MACKRL}

\newcommand{\g}{\mathcal{G}}

%% file: core/0_abstract_v2.tex
\begin{abstract}
Cooperative multi-agent reinforcement learning 
often requires decentralised policies, 
which severely limit the agents' ability 
to coordinate their behaviour.
In this paper, 
we show that common knowledge between agents 
allows for complex decentralised coordination.
Common knowledge arises naturally 
in a large number of decentralised cooperative multi-agent tasks, 
for example, when agents can reconstruct parts of each others' observations.
Since agents can independently agree on their common knowledge, they can execute complex coordinated policies that condition on this knowledge in a fully decentralised fashion.
We propose \emph{multi-agent common knowledge reinforcement learning} (\method), 
a novel stochastic actor-critic algorithm 
that learns a hierarchical policy tree.
Higher levels in the hierarchy coordinate groups of agents 
by conditioning on their common knowledge,
or delegate to lower levels with smaller subgroups
but potentially richer common knowledge.
The entire policy tree can be executed in a fully decentralised fashion.
As the lowest policy tree level consists of independent policies for each agent,
\method~reduces to independently learnt decentralised policies as a special case.
We demonstrate that  our method can exploit common knowledge 
for superior performance on complex decentralised coordination tasks, 
including a stochastic matrix game and challenging problems in StarCraft II unit micromanagement.
\end{abstract}

%% file: core/1_introduction.tex
\section{Introduction}
\label{sec:introduction}

Cooperative multi-agent problems are ubiquitous, for example, in the coordination of autonomous cars~\citep{cao2013overview} or unmanned aerial vehicles~\citep{pham_cooperative_2018,xu_multi-vehicle_2018}. %
However, how to learn control policies for such systems remains a major open question. 

\emph{Joint action learning} \citep[JAL, ][]{claus1998dynamics} learns centralised policies that select joint actions conditioned on the global state or joint observation.  
In order to execute such policies, the agents need access to either the global state or an instantaneous communication channel with sufficient bandwidth to enable them to aggregate their individual observations. These requirements often do not hold in practice, but even when they do, learning a centralised policy can be infeasible as the size of the joint action space grows exponentially in the number of agents. By contrast, \emph{independent learning} \citep[IL, ][]{tan1993multi} learns fully decentralisable policies but introduces nonstationarity as each agent treats the other agents as part of its environment.

These difficulties motivate an alternative approach: centralised training of decentralised policies.  During learning the agents can share observations, parameters, gradients, etc.\ without restriction but the result of learning is a set of decentralised policies such that each agent can select actions based only on its individual observations.  

While significant progress has been made in this direction~\citep{rashid2018qmix,foerster2016learning,foerster2017stabilising,foerster2017counterfactual,kraemer2016multi,jorge2016learning}, the requirement that policies must be fully decentralised severely limits the agents' ability to coordinate their behaviour. Often agents are forced to ignore information in their individual observations that would in principle be useful for maximising reward, because acting on it would make their behaviour less predictable to their teammates. This limitation is particularly salient in IL, which cannot solve many coordination tasks \citep{claus1998dynamics}.

\begin{wrapfigure}{R}{0.3\textwidth}
    \vspace{-9mm}
    \begin{center}
        \includegraphics[width=0.27\textwidth]{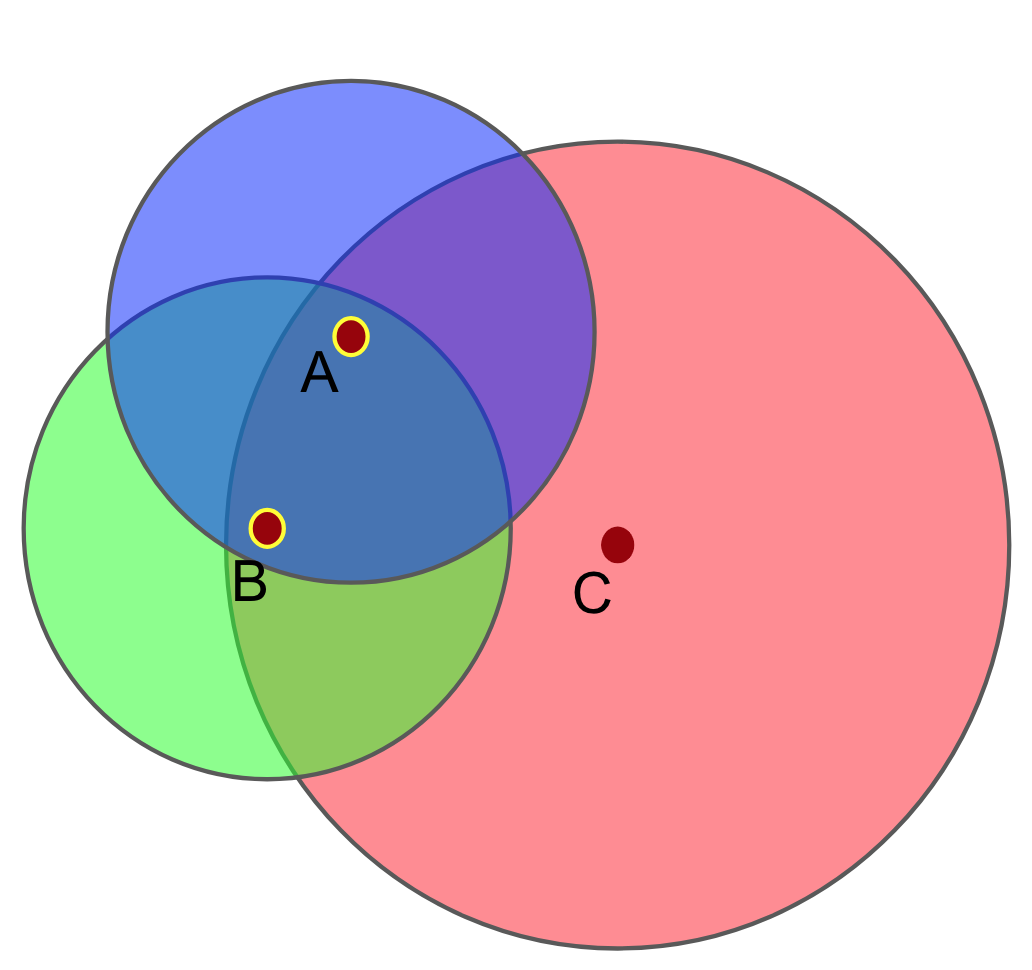}
    \end{center}
    \vspace{-4mm}
    \caption{{\footnotesize Three agents and their fields of view.  
    		A and B's locations are common knowledge to A and B 
    		as they are within each other's fields of view.  
    		Although C can see A and B, 
    		it shares no common knowledge with them.} }
    \label{fig:common_info}
    \vspace{-2mm}
\end{wrapfigure}

\textit{Common knowledge} for a group of agents consists of 
facts that all agents know and ``each individual knows that all other 
individuals know it, each individual knows that
all other individuals know that all the individuals know it, 
and so on''~\citep{osborne1994course}.
This may arise in a wide range of multi-agent problems, e.g., 
whenever a reliable communication channel is present. 
But common knowledge can also arise without communication,
if agents can infer some part of each other's observations.
For example, if each 
agent can reliably observe objects within its field of view 
and the agents know each other's fields of view, 
then they share common knowledge whenever they see each other. 
This setting is illustrated in Figure~\ref{fig:common_info}
and applies to a range of real-world scenarios, 
for example, to robo-soccer \citep{genter2017three},
fleets of self-driving cars
and multi-agent StarCraft micromanagement \citep{synnaeve2016torchcraft}.

In the absence of common knowledge, complex decentralised coordination has to rely on implicit communication, i.e., observing each other's actions or their effects \citep{heider_experimental_1944,rasouli_agreeing_2017}.  However, implicit communication protocols for complex coordination problems are difficult to learn and, as they typically require multiple timesteps to execute, can limit the agility of control during execution \citep{tian_learning_2018}. By contrast, coordination based on common knowledge is simultaneous, that is, does not require learning communication protocols \citep{halpern_knowledge_2000}.

In this paper, we introduce \emph{multi-agent common knowledge reinforcement learning} (\method), a novel stochastic policy actor-critic algorithm that can learn complex coordination policies end-to-end by exploiting common knowledge between groups of agents at the appropriate level. \method~uses a hierarchical policy tree in order to dynamically select the right level of coordination. By conditioning joint policies on common knowledge of groups of agents,  \method~occupies a unique middle ground between IL and JAL, while remaining fully decentralised.

Using a proof-of-concept matrix game, 
we show that \method~outperforms both IL and JAL. 
Furthermore, we use a noisy variant of the same matrix game 
to show that \method~can exploit a weaker form of group knowledge called 
\textit{probabilistic common knowledge} \citep{krasucki_probabilistic_1991}, 
that is induced by agent beliefs over common knowledge, 
derived from noisy observations. 
We show that \method's performance degrades gracefully with increasing noise levels.

We then apply \method~to challenging StarCraft II 
unit micromanagement tasks \citep{Vinyals2017} 
from the {\em StarCraft Multi-Agent Challenge} \cite[SMAC,][]{smac2018pre}. 
We show that simultaneous coordination based on pairwise common knowledge 
enables \method~to outperform the state-of-the-art algorithms 
COMA \citep{foerster2017counterfactual} and QMIX \citep{rashid2018qmix} 
and provide a variant of \method~that scales to tasks with many agents.

%% file: core/3_background.tex
\section{Problem Setting}
\label{sec:problem}
\newcommand{\envstate}{s^\text{env}}

\vspace{-2mm}
Cooperative multi-agent tasks with $n$ agents $a \in \Set A$ can be modelled as 
\emph{decentralised partially observable Markov decision processes} 
\citep[Dec-POMDPs,][]{Oliehoek08JAIR}. 
The state of the system is $s \in \Set S$. At each time-step, each agent $a$
receives an observation $z^a \in \Set Z$  and can select an action 
$u^a_\text{env} \in \Set U^a_\text{env}$.
We use the $\text{env}$-subscript to denote actions executed by the agents in 
the environment, as opposed to latent `actions' that may be taken by 
higher-level controllers of the hierarchical method introduced in Section \ref{sec:method}.
Given a joint action $\vu_\text{env} := (u^1_\text{env}, \ldots, u^n_\text{env}) \in \Set U_\text{env}$,
the discrete-time system dynamics draw the successive state
$s' \in \Set S$ from the conditional distribution $P(s'|s, \vu_\text{env})$
and yield a cooperative reward according to the function $r(s, \vu_\text{env})$.

The agents aim to maximise the discounted return 
$R_t = \sum_{l=0}^H \gamma^l \, r(s_{t+l}, \vu_{t+l, \text{env}})$
from episodes of length $H$.
The joint policy $\pi(\vu_\text{env}|s)$ is restricted
to a set of decentralised policies $\pi^a(u_\text{env}^a|\tau^a_t)$
that can be executed independently, i.e., each agent's policy conditions only on 
its own action-observation history 
$\tau^a_t := [z^a_0, u^a_0, z^a_1, \ldots, z^a_t]$.
Following previous work \citep{rashid2018qmix,foerster2016learning,%
foerster2017stabilising,foerster2017counterfactual,kraemer2016multi,jorge2016learning},
we allow decentralised policies to be learnt in a centralised fashion.

\textbf{Common knowledge} of a group of agents $\g$
refers to facts that all members know, 
and that ``each individual knows that all other individuals know it,  
each individual knows that all other individuals know that all 
the individuals know it, and so on'' \cite{osborne1994course}.
Any data $\xi$ that are known to all agents before execution/training, 
like a shared random seed, are obviously common knowledge.
Crucially, every agent $a \in \g$ can deduce 
the same {\em history of common knowledge} $\tau^\g_t$
from its own history $\tau^a_t$ and the commonly known data $\xi$,
that is, $\tau^\g_t := \mathcal{I}^{\Set G}{(\tau^a_t, \xi)} 
= \mathcal{I}^{\Set G}{(\tau^{\bar a}_t, \xi)}, \forall a, \bar a \in \g$. 
Furthermore, any actions taken by 
a policy $\pi^\g(\mathbf{u}^\g_\text{env}|\tau^\g_t)$
over the group's joint action space $\mathcal{U}^\g_{\text{env}}$
are themselves common knowledge,
if the policy is deterministic or pseudo-random with a shared random seed
and conditions only on the common history $\tau^\g_t$,
i.e. the set formed by restricting each transition tuple within the joint history of agents in $\g$ to what is commonly known in $\g$ at time $t$.
Common knowledge of subgroups $\g' \subset \g$ cannot decrease,
that is, $\Set I^{\g'}(\tau^a_t, \xi) \supseteq \Set I^{\g}(\tau^a_t, \xi)$.

Given a Dec-POMDP with noisy observations, 
agents in a group $\g$ might not be able to establish 
true common knowledge even if sensor noise properties are commonly known \citep{halpern_knowledge_2000}. 
Instead, each agent $a$ can only deduce 
its own \textit{beliefs} $\tilde{\mathcal{I}}^{\Set G}_a{(\tilde{\tau}^a_t)}$ 
over what is commonly known within $\mathcal{G}$, where $\tilde{\tau}^a_t$ 
is the agent's belief over what constitutes the groups' common history.
Each agent $a$ can then evaluate its own \textit{belief over the group policy} $\tilde{\pi}^\g_a(u^\g_\text{env}|\tilde{\tau}^\g_t)$.  In order to minimize the probability of disagreement during decentralized group action selection, agents in $\g$ can perform \textit{optimal correlated sampling} based on a shared random seed \citep{holenstein_parallel_2007,bavarian_optimality_2016}.
For a formal definition of \textit{probabilistic common knowledge}, please refer to
\citet[Definitions 8 and 13]{krasucki_probabilistic_1991}.

\textbf{Learning under common knowledge (LuCK)} 
is a novel cooperative multi-agent reinforcement learning setting, 
where a Dec-POMDP is augmented 
by a common knowledge function $\mathcal{I}^\mathcal{G}$ 
(or \textit{probabilistic} common knowledge function $\tilde{\mathcal{I}}_a^\mathcal{G}$).
Groups of agents $\g$ can coordinate by learning policies 
that condition on their common knowledge.
In this paper $\mathcal{I}^\mathcal{G}$ (or $\tilde{\mathcal{I}}_a^\mathcal{G}$)
is fixed apriori, 
but it could also be learnt during training. 
The setting accommodates a wide range of 
real-world and simulated multi-agent tasks. 
Whenever a task is cooperative and learning is centralised, 
then agents can naturally learn suitable 
$\mathcal{I}^{\mathcal{G}}$ or $\tilde{\mathcal{I}}_a^{\mathcal{G}}$. 
Policy parameters can be exchanged during training as well
and thus become part of the commonly known data $\xi$.
Joint policies  
where coordinated decisions of a group $\g$ 
only condition on the common knowledge of $\g$
can be executed in a fully decentralised fashion. 
In Section \ref{sec:method} we introduce \method, 
which uses centralised training to 
learn fully decentralised policies under common knowledge.

\textbf{Field-of-view common knowledge} is a form of 
\textit{complete-history common knowledge} \citep{halpern_knowledge_2000}, 
that arises within a Dec-POMDP 
if agents can deduce parts of other agents' observations from their own.
In this case, 
an agent group's common knowledge is the intersection of 
observations that all members can reconstruct from each other.
In Appendix \ref{sec:common_knowledge} we formalise this concept 
and show that, under some assumptions, 
common knowledge is the intersection 
of all agents' sets of visible objects, 
if and only if all agents can see each other.
Figure \ref{fig:common_info} shows an example 
for three agents with circular fields of view.  
If observations are noisy, 
each agent bases its belief on its own noisy observations 
thus inducing an equivalent form of probabilistic common knowledge
$\tilde{\mathcal{I}}_a^\mathcal{G}$.

Field-of-view common knowledge naturally 
occurs in many interesting real-world tasks, 
such as autonomous driving \citep{cao2013overview} 
and robo-soccer \citep{genter2017three}, 
as well as in simulated benchmarks such as StarCraft II \citep{Vinyals2017}. 
A large number of cooperative multi-agent tasks 
can therefore benefit from common knowledge-based coordination 
introduced in this paper.

%% file: core/4_method_v2.tex
\section{Multi-Agent Common Knowledge Reinforcement Learning}
\label{sec:method}
\def\env{\text{env}}

The key idea behind \method~is to learn decentralised policies 
that are nonetheless coordinated by common knowledge.
As the common knowledge history $\tau^\g_t$ of a group of agents $\g$
can be deduced by every member, 
i.e., $\tau^\g_t = \Set I^\g(\tau^a_t, \xi), \forall a \in \g$,
any deterministic function based only on $\tau^\g_t$ 
can thus be independently computed by every member as well.
The same holds for pseudo-random functions like stochastic policies, 
if they condition on a commonly known random seed in $\xi$.

\method~uses a hierarchical policy 
$\pi(\vu_\env|\{\tau^a_t\}_{a \in \Set A}, \xi)$
over the joint environmental action space of all agents $\Set U_\env$.
The hierarchy forms a tree of sub-policies $\pi^\g$ over groups $\g$,
where the root $\pi^{\Set A}$ covers all agents.
Each sub-policy $\pi^\g(u^\g|\,\Set I^\g(\tau^{\g}_t, \xi))$ 
conditions on the common knowledge of $\g$, 
including a shared random seed in $\xi$,
and can thus be executed by every member of $\g$ independently.
The corresponding action space $\Set U^\g$
contains the environmental actions 
of the group $\vu^\g_\env \in \Set U^\g_\env$
{\em and/or} a set of group partitions, 
that is, $\vu^\g = \{\g_1, \ldots, \g_k\}$ 
with $\g_i \cap \g_j = \varnothing, \;
\forall i \neq j$
and $\cup_{i=1}^k \g_i = \g$.

Choosing a partition $\vu^\g \neq \Set U^\g_\env$ 
yields control to the sub-policies $\pi^{\g_i}$ 
of the partition's subgroups $\g_i \in \vu^\g$.
This can be an advantage in states 
where the common history $\tau^{\g_i}_t$ %
of the subgroups is more informative than $\tau^\g_t$. %
All action spaces have to be specified in advance,
which induces the hierarchical tree structure of the joint policy.
Algorithm \ref{alg:decentral} shows the decentralised
sampling of environmental actions from the hierarchical joint policy
as seen by an individual agent $a \in \Set A$.

\input{core/pairwise_figure}
\input{core/algorithm_decentralised}

As the common knowledge of a group with only one agent $\g=\{a\}$ is 
$\Set I^{\{a\}}(\tau^a,\xi) =\tau^a$, fully decentralised policies are a special
case of \method~policies: in this case, the root policy $\pi^{\Set A}$
has only one action $\Set U^{\Set A} := \{\vu^{\Set A}\}$,
$\vu^{\Set A} := \{ \{1\}, \ldots, \{n\} \}$,
and all leaf policies $\pi^{\{a\}}$ 
have only environmental actions $\Set U^{\{a\}} := \Set U^a_\env$.

\subsection{Pairwise \method}
\label{sec:pairwise_mackrl}
To give an example of one possible \method~architecture,
we define {\em Pairwise \method}, illustrated in Figure~\ref{fig:mackrl}.
As joint action spaces grow exponentially in the number of agents,
we restrict ourselves to \textit{pairwise} joint policies and
define a three-level hierarchy of controllers.  

The root of this hierarchy is the \emph{pair selector} $\pi^{\Set A}_\text{ps}$,
with an action set $\Set U^{\Set A}_\text{ps}$ that contains
all possible partitions of agents into pairs 
$\{ \{a_1, \bar a_1\}, \ldots,  \{a_{n/2}, \bar a_{n/2}\}\} =: \vu^{\Set A} 
\in \Set U^{\Set A}_\text{ps}$, but no environmental actions.
If there are an odd number of agents, 
then one agent is put in a singleton group.
At the second level, each \emph{pair controller}  $\pi^{a \bar a}_\text{pc}$ 
of the pair $\g=\{a,\bar a\}$
can choose between joint actions 
$\vu^{a \bar a}_\env \in \Set U^{a}_\env \times \Set U^{\bar a}_\env$ 
and one delegation action $u_d^{a \bar a} := \{\{a\},\{\bar a\}\}$, 
i.e., $\mathcal{U}^{a \bar a}_\text{pc} := \mathcal{U}^a_\text{env} 
\times \mathcal{U}^{\bar a}_\text{env} \union \{u_d^{a \bar a}\}$.
At the third level, individual controllers $\pi^a$ select an
individual action $u^a_\text{env} \in \Set U^a_{\text{env}}$ 
for a single agent $a$. 
This architecture retains manageable joint action spaces,
while considering all possible pairwise coordination configurations.
Fully decentralised policies are the special case 
when all pair controllers always choose 
partition $u_d^{a \bar a}$ to delegate.

Unfortunately, 
the number of possible pairwise partitions is $O(n!)$,
which limits the algorithm to medium sized sets of agents.
For example, $n=11$ agents induce 
$|\Set U^{\Set A}_\text{ps}| = 10395$ unique partitions.
To scale our approach to tasks with many agents, 
we share network parameters between all pair controllers
with identical action spaces, thereby greatly improving sample efficiency.
We also investigate a more scalable variant in which the action space of 
the pair selector $\pi^{\Set A}_\text{ps}$ is only a fixed random subset
of all possible pairwise partitions.  
This restricts agent coordination to a smaller set of predefined pairs, 
but only modestly affects \method's performance (see Section \ref{sec:exp_sc2}).

\subsection{Training}
\input{core/algorithm_centralised}

The training of policies in the \method~family is based on 
{\em Central-V} \citep{foerster2017counterfactual},
a stochastic policy gradient algorithm \citep{williams1992simple}
with a centralised critic.
Unlike the decentralised policy, 
we condition the centralised critic on the state $s_t \in \Set S$
and the last actions of all agents $\vu_{\env, t-1} \in \Set U_\env$.
We do not use the multi-agent counterfactual baseline 
proposed by \citet{foerster2017counterfactual}, 
because \method~effectively turns training into a single agent problem 
by inducing a correlated  probability across the joint action space.
Algorithm \ref{alg:central} shows how the 
probability of choosing a joint environmental action 
$\vu^\g_\env \in \Set U^\g_\env$ of group $\g$ is computed: 
the probability of choosing the action in question
is added to the recursive probabilities 
that each partition $\vu^\g \not\in \Set U^\g_\env$
would have selected it.
During decentralized execution,
Algorithm \ref{alg:decentral} only traverses 
one branch of the tree.
To further improve performance, 
all policies choose actions greedily outside of training 
and do thus not require any additional means 
of coordination such as shared random seeds during execution.

At time $t$, the gradient with respect to the parameters $\theta$ 
of the joint policy $\pi(\vu_\env | \{\tau^a_t\}_{a \in \Set A}, \xi)$ 
is:
\begin{equation}
	\nabla_{\!\theta} J_t = 
	\big(\! \underbrace{ r(s_{t}, \vu_{\text{env},t}) 
		+ \gamma V(s_{t+1}, \vu_{\text{env},t})  
		-  V(s_{t}, \vu_{\text{env},t-1}) 
		}_{\text{sample estimate of the advantage function}}
	\!\big) \;
	\nabla_{\!\theta}  \log\!\big(\!\underbrace{
		\pi(\vu_{\text{env}, t} | 
			\{\tau^a_t\}_{a \in \Set A}, \xi) 
	}_{\hspace{-3cm}\text{\sc joint\_policy}(\vu_{\env,t} | 
		\Set A, \{\tau^a_t\}_{a \in \Set A}, \xi) \hspace{-2cm}
	} \!\big) ,
\end{equation}

The value function $V$ is learned by gradient descent 
on the TD($\lambda$) loss \citep{sutton1998reinforcement}.

As the hierarchical \method~policy tree computed 
by Algorithm \ref{alg:central} is fully differentiable 
and \method~trains a joint policy in a centralised fashion, 
the standard convergence results 
for actor-critic algorithms \citep{konda1999actor}
with compatible critics \citep{sutton1999policy}
apply.

%% file: core/pairwise_figure.tex
\begin{figure*}%
    \centering
    \includegraphics[width=0.35\textwidth]{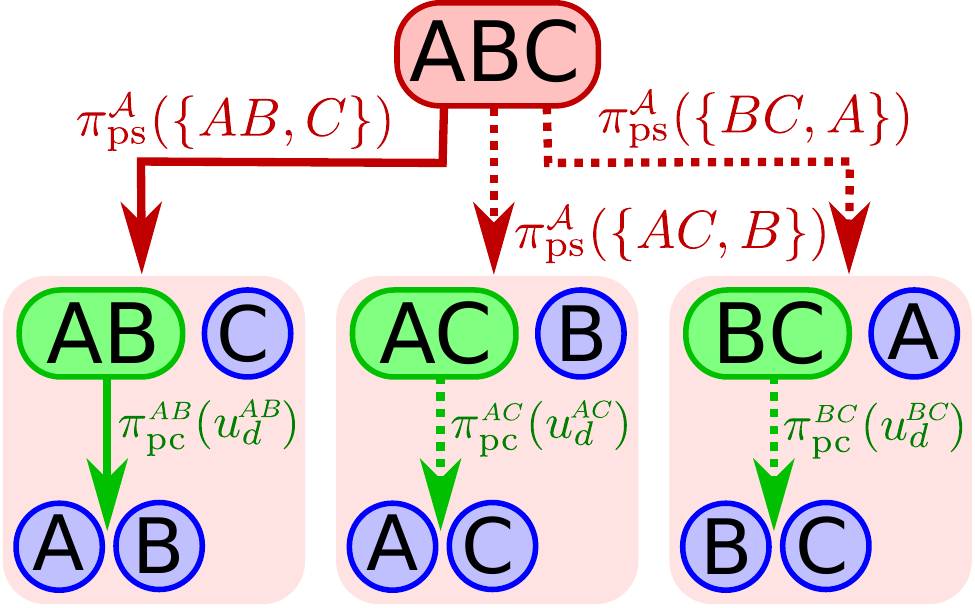}
    \hfill
	\includegraphics[width=0.64\textwidth]{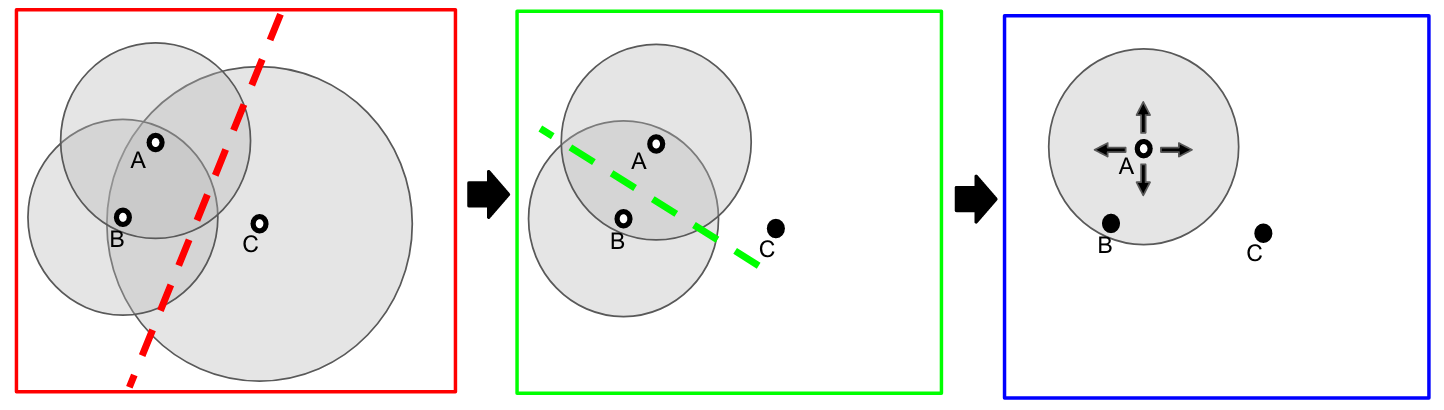}
	\vspace{-4mm}
	\captionsetup{singlelinecheck=off}
 	\caption[foo bar]{ \small
 		An illustration of Pairwise \method.
		[left]: 
		the full hierarchy for 3 agents 
		(dependencies on common knowledge are omitted for clarity).
		Only solid arrows are computed during decentralised sampling 
		with Algorithm \ref{alg:decentral},
		while all arrows must be computed recursively
		during centralised training (see Algorithm \ref{alg:central}). 
 		[right]: 
 		the (maximally) 3 steps of decentralised sampling 
 		from the perspective of agent A. 
 		\begin{enumerate}
 		\item Pair selector $\pi_\text{ps}^{\Set A}$ 
 		chooses the partition $\{AB,C\}$
 		based on the common knowledge of all agents 
 		$\mathcal{I}^{ABC}{(\tau^A, \xi)} = \varnothing$.
 		\item Based on the common knowledge of pair $A$ and $B$,
 		$\mathcal{I}^{AB}{(\tau^A, \xi)}$, 
 		the pair controller $\pi_\text{pc}^{AB}$ 
 		can either choose a joint action 
 		$(u_\text{env}^A,u_\text{env}^B)$, 
 		or delegate to individual controllers by selecting $u_d^{AB}$.
 		\item If delegating, the individual controller $\pi^A$
 		must select the action $u_\text{env}^A$ for the single agent $A$.
 		All steps can be computed based on A's history $\tau^A$.
 		\end{enumerate}
	 }
    \label{fig:mackrl}
    \vspace{-4mm}
\end{figure*}

%% file: core/algorithm_decentralised.tex
\begin{algorithm}[b!]
\begin{algorithmic}
	\Function{select\_action}{$a, \tau^a_{t}, \xi$}
			\Comment{random seed in $\xi$ is common knowledge}
		\State $\g := \mathcal{A}$
			\Comment{initialise the group $\g$ of all agents}
		\State $\vu^\g_t \sim \pi^{\g}\big(\cdot |\, 
					\mathcal{I}^{\g}(\tau^a_t, \xi) \big)$
			\Comment{$\vu^\g_t$ is either a joint environmental 
						action in $\mathcal{U}^\g_\text{env}$...}
		\While{$\vu^\g_t \not\in \mathcal{U}^\g_\text{env}$}  
				\Comment{... or a set of disjoint subgroups $\{\g_1, \ldots, \g_k\}$}
			\State $\g := \{\g' \,|\, a \in \g', \g' \in \vu^\g_t\}$
				\Comment{select subgroup containing agent $a$}
			\State $\vu^\g_t \sim \pi^{\g}\big(\cdot |\, 
					\mathcal{I}^{\g}(\tau^a_t, \xi) \big)$
				\Comment{draw an action for that subgroup}
			\vspace{1mm}
		\EndWhile	
		\Return $u^a_t$
			\Comment{return environmental action 
					$u^a_t \in \Set U^a_\text{env}$ of agent $a$}
	\EndFunction
\end{algorithmic}
\caption{Decentralised action selection for agent $a\in\mathcal{A}$ in MACKRL} 
\label{alg:decentral}
\end{algorithm}

%% file: core/algorithm_centralised.tex
\begin{algorithm}[t!]
\def\funname{joint\_policy}
\begin{algorithmic}
	\Function{\funname}{$\vu^\g_\text{env} |\, \Set G, 
				\{\tau^a_{t}\}_{a \in \Set G}, \xi$}
			\Comment{random seed in $\xi$ is common knowledge}
		\vspace{.5mm}
		\State $a' \sim \Set G \,; \quad \vI^\g := \Set I^\g(\tau^{a'}_t, \xi)$ 
				\Comment{common knowledge $\vI^\g$ is identical 
						for every agent $a' \in \g$}
		\vspace{.5mm}		
		\State $p_\text{env} := 0$
			\Comment{initialise probability for choosing 
					environmental joint action $\vu^\g_\text{env}$}
		\vspace{.5mm}
		\For{$\vu^\g \in \Set U^\g$}
				\Comment{add probability to choose $\vu^\g_\text{env}$
						for all outcomes of $\pi^\g$ }
			\If{$\vu^\g = \vu^\g_\text{env}$}
					\Comment{if $\vu^\g$ is the environmental joint action in question}
				\State$p_\text{env} := p_\text{env} +
					\pi^\g\!\big(\vu^\g_\text{env} |\, \vI^\g \big)$ 
			\EndIf
			\vspace{.5mm}
			\If{$\vu^\g \not\in \Set U^\g_\text{env}$}
					\Comment{if $\vu^\g = \{\g^1, \ldots, \g^k\}$ 
						is a set of disjoint subgroups}
				\State $p_\text{env} := p_\text{env} + 
					\pi^\g\!\big(\vu^\g| \,\vI^\g \big) 
					{\prod\limits_{\g' \in \vu^\g}} \text{\sc \funname}%
						\big(\vu^{\g'}_\text{env} |\, \g', 
						\{\tau_t^a\}_{a \in \g'}, \xi \big)$
			\EndIf
		\EndFor
		\Return $p_\text{env}$
			\Comment{return probability that controller $\pi^\g$ 
					would have chosen $\vu^\g_\text{env}$}
	\EndFunction
\end{algorithmic}
\caption{Compute joint policies for a given 
		$\vu^\g_\text{env} \in \Set U^\g_\text{env}$ 
		of a group of agents $\g$ in MACKRL} 
\label{alg:central}
\end{algorithm}

%% file: core/6_experiments.tex
\section{Experiments and Results}
\label{sec:results}

We evaluate Pairwise \method~(henceforth referred to as \method) on two environments\footnote{All source code is available at \url{https://github.com/schroederdewitt/mackrl}.\vspace{4mm}} : first, we use a matrix game with special coordination requirements to illustrate \method's ability to surpass both IL and JAL. Secondly, we 
employ \method~with deep recurrent neural network policies in order to outperform state-of-the-art baselines
on a number of challenging StarCraft II unit micromanagement tasks. 
Finally, we demonstrate \method's robustness to sensor noise and its scalability to large numbers of agents to illustrate its applicability to real-world tasks.

\subsection{Single-step matrix game}
\label{sec:matrix_results}

To demonstrate how \method~trades off between 
independent and joint action selection, 
we evaluate a two-agent matrix game with partial observability. 
In each round, a fair coin toss decides which of the two matrix games 
in Figure \ref{fig:matrix_games} [left] is played.
Both agents can observe which game has been selected 
if an observable \emph{common knowledge bit} is set.
If the bit is not set, 
each agent observes the correct game only with probability $p_\sigma$,
and is given \emph{no observation} otherwise. 
Crucially, whether agents can observe the current game 
is in this case determined independently of each other.
Even if both agents can observe which game is played,
this observation is no longer common knowledge
and cannot be used to infer the choices of the other agent.
To compare various methods 
we adjusted $p_\sigma$ such that the 
independent probability of each agent 
to observe the current game is fixed at 75\%.

\input{core/matrix_figure}

In order to illustrate \method's performance, we compare it to three other methods: Independent Actor Critic \citep[IAC,][]{foerster2017counterfactual} is a variant of Independent Learning where each agent conditions both its decentralized actor and critic only on its own observation. Joint Action Learning \citep[JAL,][]{claus1998dynamics} learns a centralized joint policy that conditions on the union of both agent's observations. CK-JAL is a decentralised variant of JAL in which both agents follow a joint policy that conditions only on the common knowledge available.

Figure \ref{fig:matrix_games} [middle] plots \method's performance 
relative to IL, CK-JAL, and JAL 
against the fraction of observed games that are caused by a set CK-bit.
As expected, the performance of CK-JAL linearly increases as more common knowledge becomes available, whereas the performance of IAC remains invariant. \method's performance matches the one of IAC if no common knowledge is available and matches those of JAL and CK-JAL in the limit of all observed games containing common knowledge. In the regime between these extremes, \method~outperforms both IAC and CK-JAL, but is itself upper-bounded by JAL, which gains the advantage due to central execution.

To assess \method's performance in the case of \textit{probabilistic common knowledge} (see Section \ref{sec:problem}), we also consider the case where the observed
{\em common knowledge bit} of individual agents is randomly flipped 
with probability $p$. This implies that both agents do not share true common knowledge with respect to the game matrix played. Instead, each agent $a$ can only form a belief $\tilde{\mathcal{I}}_a^\mathcal{G}$ over what is commonly known. The commonly known pair controller policy can then be conditioned on each agent's belief, resulting in agent-specific pair controller policies $\tilde{\pi}^{aa'}_{pc,a}$. 

As $\tilde{\pi}^{aa'}_{pc,a}$ and $\tilde{\pi}^{aa'}_{pc,a'}$ are no longer guaranteed to be consistent, agents need to sample from their respective pair controller policies in a way that minimizes the probability that their outcomes disagree in order to maximise their ability to coordinate. Using their access to a shared source of randomness $\xi$, the agents can optimally solve this \textit{correlated sampling} problem using Holenstein's strategy (see Appendix \ref{sec:holenstein}). However, this strategy requires the evaluation of a significantly larger set of actions and quickly becomes computationally expensive.
Instead, we use a suboptimal heuristic that nevertheless performs satisfactorily in practice and can be trivially extended to groups of more than two agents: given a shared uniformly drawn random variable $\delta\sim\xi$, $0 \leq \delta < 1$,  each agent $a$ samples an action $u_a$ such that  \vspace{-0.4em}
\begin{equation}
\sum_{u=1}^{u_a-1}\tilde{\pi}^{aa'}_{pc,a}(u)\leq\delta<\sum_{u=1}^{u_a}\tilde{\pi}^{aa'}_{pc,a}(u).\vspace{-0.4em}
\end{equation}

Figure \ref{fig:matrix_games} [right] shows that
\method's performance declines remarkably gracefully with increasing observation noise. 
Note that as real-world sensor observations tend to tightly correlate with the true observations, noise levels of $p\geq 0.1$ in the context of the single-step matrix game are rather extreme in comparison, as they indicate a completely different game matrix.
This illustrates \method's applicability to real-world tasks with noisy observation sensors.

\input{core/sc2_figure}

\subsection{StarCraft II micromanagement}
\label{sec:exp_sc2}
To demonstrate \method's ability to solve complex coordination tasks, we evaluate it on a challenging multi-agent version of StarCraft II (SCII) micromanagement. To this end, we report performance on three challenging coordination tasks from the established multi-agent benchmark SMAC \citep{smac2018pre}.

The first task, map \textit{2s3z}, contains mixed unit types, where both the \method~agent and the game engine each control two Stalkers and three Zealots. Stalkers are ranged-attack units that take heavy damage from melee-type Zealots. Consequently, a winning strategy needs to be able to dynamically coordinate between letting one's own Zealots attack enemy Stalkers, and when to backtrack in order to defend one's own Stalkers against enemy Zealots. The challenge of this coordination task results in a particularly poor performance of Independent Learning \citep{smac2018pre}.

The second task, map \textit{3m}, presents both sides with three Marines, which are medium-ranged infantry units. The coordination challenge on this map is to reduce enemy fire power as quickly as possible by focusing unit fire to defeat each enemy unit in turn. 
The third task, map \textit{8m}, scales this task up to eight Marines on both sides. The relatively large number of agents involved poses additional scalability challenges. 

On all maps, the units are subject to partial observability constraints and have a circular field of view with fixed radius. Common knowledge $\mathcal{I}^{\mathcal{G}}$ between groups $\mathcal{G}$ of agents arises through entity-based field-of-view common knowledge 
(see Section \ref{sec:problem} and Appendix \ref{sec:common_knowledge}). 

We compare \method~to Central-V \citep{foerster2017counterfactual}, 
as well as COMA \citep{foerster2017counterfactual} and QMIX \citep{rashid2018qmix}, 
where the latter is an off-policy value-based algorithm 
that is the current state-of-the-art on all maps. 
We omit IL results since 
it is known to do comparatively poorly ~\citep{smac2018pre}. 
All experiments use SMAC settings for comparability 
(see \citet{smac2018pre}  and Appendix \ref{sec:sc2_exp_supp} for details). 
In addition, \method~ and its within-class baseline Central-V 
share equal hyper-parameters as far as applicable.

\begin{wrapfigure}{R}{.36\linewidth}
	\vspace{-4mm}
	\includegraphics[width=\linewidth]{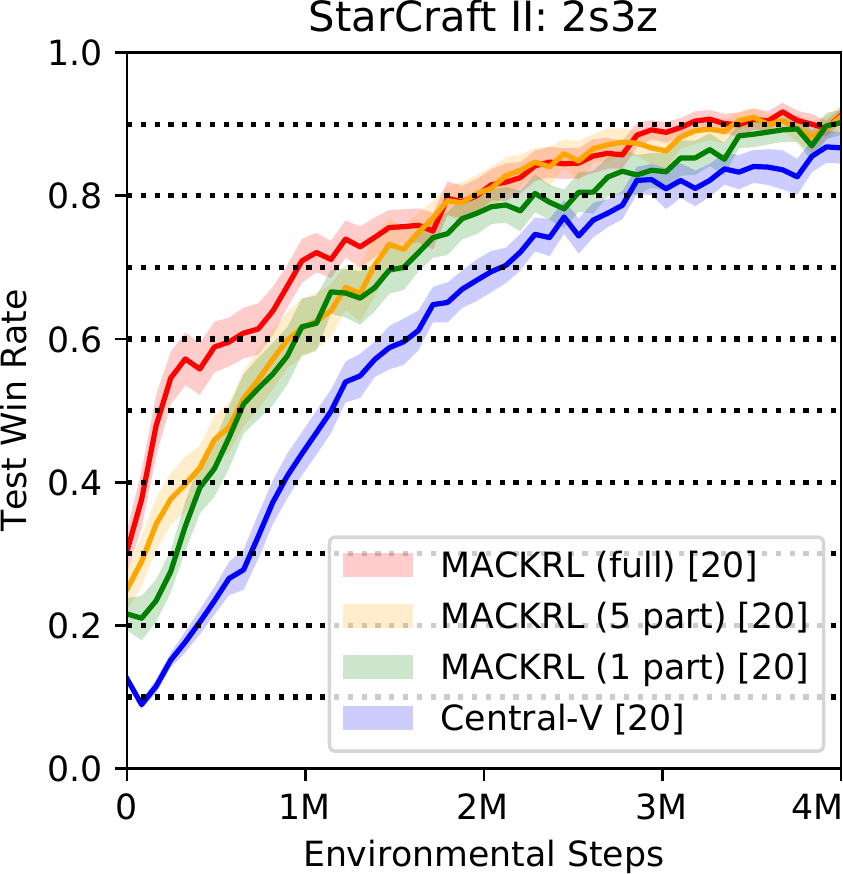}
	\vspace{-6mm}
	\caption{\small Illustrating \method's scalability properties using partition subsamples of different sizes.}
	\label{fig:scalability}
	\vspace{-4mm}
\end{wrapfigure}

\method~outperforms the Central-V baseline 
in terms of sample efficiency and limit performance on all maps 
(see Figure \ref{fig:res_sc2}).
All other parameters being equal, 
this suggests that \method's superiority over Central-V 
is due to its ability to exploit common knowledge. 
In Appendix \ref{sec:pair_intro}, we confirm this conclusion by showing that the policies learnt by the pair controllers are almost always preferred over individual controllers whenever agents have access to substantial amounts of common knowledge. 

\method~also significantly outperforms COMA and QMIX 
on all maps in terms of sample efficiency, 
with a similar limit performance to QMIX (see Figure \ref{fig:res_sc2}). 
These results are particularly noteworthy as \method~employs 
neither a sophisticated multi-agent baseline, like COMA, 
nor an off-policy replay buffer, like QMIX. 

As mentioned in Section \ref{sec:pairwise_mackrl}, the number of 
possible agent partitions available to the pair selector $\pi^{\Set A}_\text{ps}$ 
grows as $\Set O(n!)$.
We evaluate a scalable variant of \method~that constrains the number of partitions
to a fixed subset, which is drawn randomly before training. 
Figure \ref{fig:scalability} shows that sample efficiency 
declines gracefully with subsample size. 
\method's policies appear able to exploit 
any common knowledge configurations available, 
even if the set of allowed partitions is not exhaustive.

%% file: core/matrix_figure.tex
\begin{figure}[!t]
  \vspace{-2mm}
   \begin{minipage}{.24\columnwidth}\footnotesize%
   \vspace{-1.0em}
   \[{\displaystyle
   \frac{1}{5}
   \begin{bmatrix}
      5 & {\scriptstyle 0} & {\scriptstyle 0} & 2 & {\scriptstyle 0} \\
      {\scriptstyle 0} & 1 & 2 & 4 & 2 \\
      {\scriptstyle 0} & {\scriptstyle 0} & {\scriptstyle 0} & 2 & {\scriptstyle 0} \\
      {\scriptstyle 0} & {\scriptstyle 0} & {\scriptstyle 0} & 1 & {\scriptstyle 0} \\
      {\scriptstyle 0} & {\scriptstyle 0} & {\scriptstyle 0} & {\scriptstyle 0} & 5 \\
   \end{bmatrix}   
   \vspace{-2.5em}}
   \]\\\vspace{-1.0em}
    \[
   {\displaystyle
   \frac{1}{5}
   \begin{bmatrix}
      {\scriptstyle 0} & {\scriptstyle 0} & 1 & {\scriptstyle 0} & 5 \\
      {\scriptstyle 0} & {\scriptstyle 0} & 2 & {\scriptstyle 0} & {\scriptstyle 0} \\
      1 &  2 & 4  & 2 & 1 \\
      {\scriptstyle 0} & {\scriptstyle 0} & 2 & {\scriptstyle 0} & {\scriptstyle 0} \\
      5 & {\scriptstyle 0} & 1 & {\scriptstyle 0} & {\scriptstyle 0} \\
   \end{bmatrix}
   \vspace{-2.5em}}
   \]  
  \end{minipage}
  \begin{minipage}{0.365\columnwidth} \small
    \begin{center}
        \includegraphics[height=1.85in]{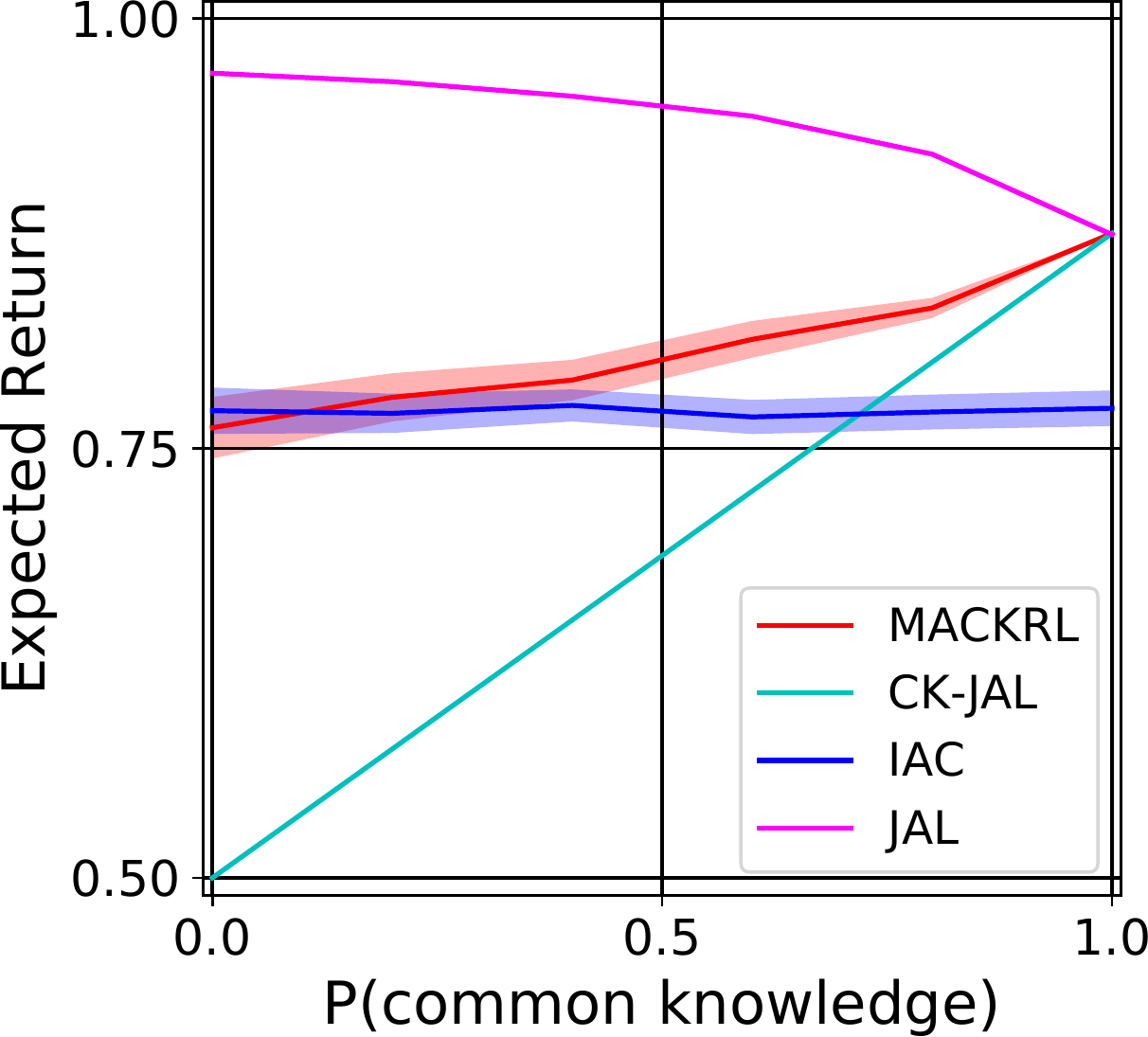}
    \end{center}
  \end{minipage}
   \begin{minipage}{0.365\columnwidth} \small
    \begin{center}
        \includegraphics[height=1.85in]{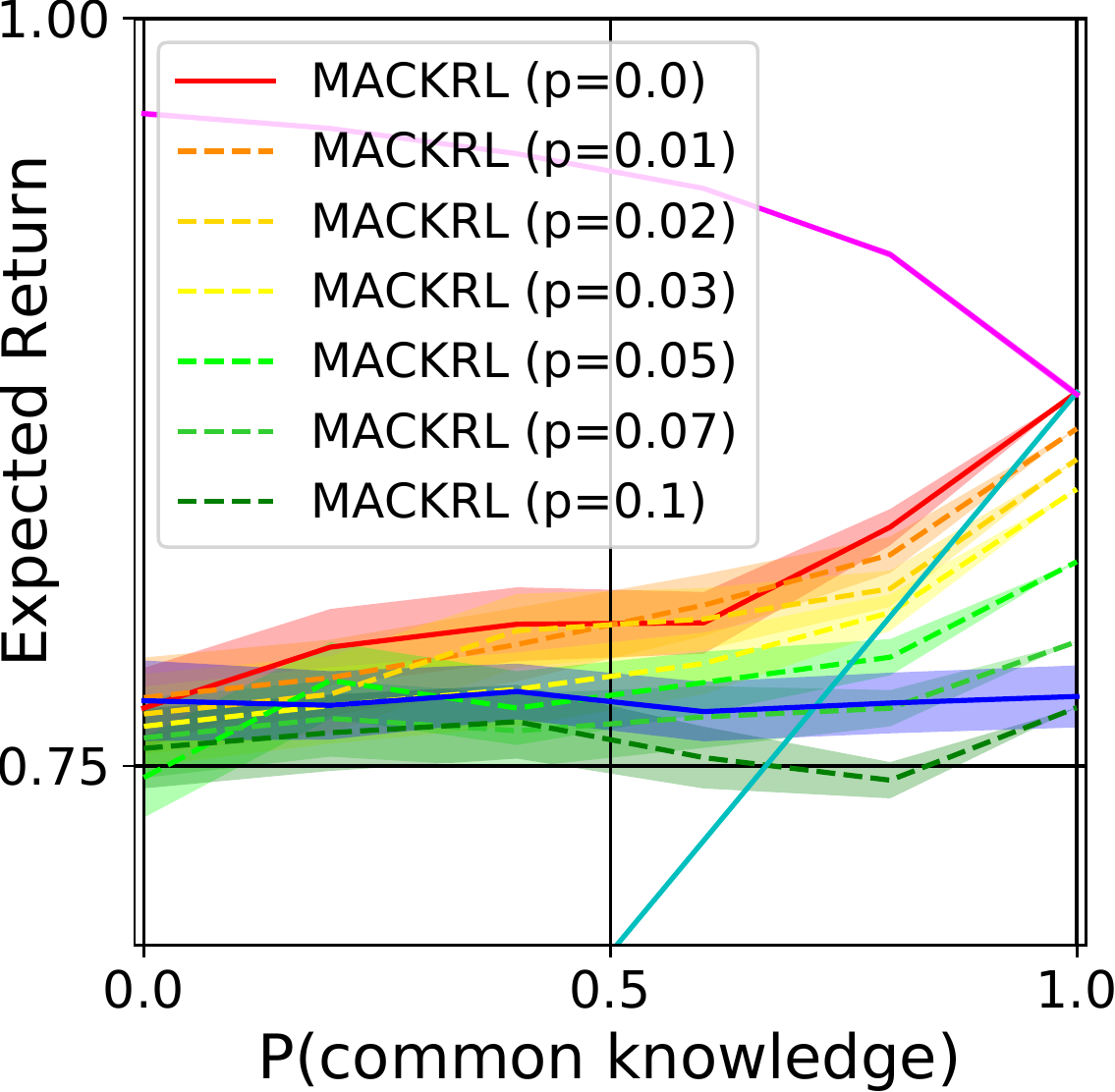}
    \end{center}
  \end{minipage}
  \caption{Game matrices A (top) and B (bottom) [left]. \method~almost always outperforms both IL and CK-JAL and is upper-bounded by JAL [middle]. When the common knowledge is noised by randomly flipping the CK-bit with probability $p$, \method~degrades gracefully [right].}
  \label{fig:matrix_games}
  \vspace{-2mm}
\end{figure}

%% file: core/sc2_figure.tex
\begin{figure*}[b!]
	\begin{center}
      \begin{center}
         \includegraphics[width=\columnwidth]{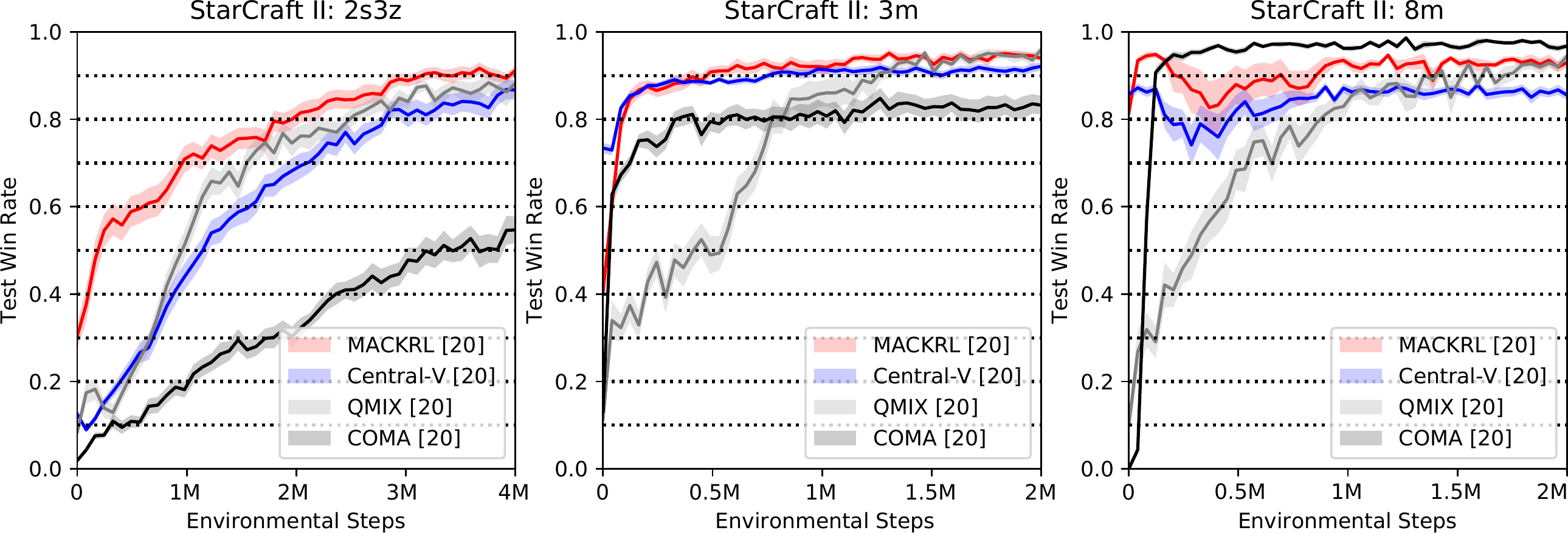}
      \end{center}
	\vspace{-2mm}
	\caption{ \small
		Win rate at test time across StarCraft II scenarios:
		2 Stalkers \& 3 Zealots [left], 3 Marines [middle] and	8 Marines [right]. 
		Plots show means and their standard errors with [number of runs]. 
	} 	\label{fig:res_sc2}
	\end{center}
\end{figure*}

%% file: core/2_related_work.tex
\section{Related Work}
\label{sec:related_work}

\emph{Multi-agent reinforcement learning} (MARL) has been studied extensively in small environments \citep{busoniu2008comprehensive, yang2004multiagent}, but scaling it to large state spaces or many agents has proved problematic. %
\citet{guestrin2002multiagent} propose the use of \emph{coordination graphs}, which exploit conditional independence properties between agents that are captured in an undirected graphical model, in order to efficiently select joint actions. \emph{Sparse cooperative $Q$-learning} \citep{kok2004sparse} also uses coordination graphs to efficiently maximise over joint actions in the $Q$-learning update rule. Whilst these approaches allow agents to coordinate optimally, they require the coordination graph to be known and for the agents to either observe the global state or to be able to freely communicate. In addition, in the worst case there is no conditional independence to exploit and maximisation must still be performed over an intractably large joint action space. 

There has been much work on scaling MARL to handle complex, high dimensional 
state and action spaces. In the setting of fully centralised training and 
execution, \citet{usunier2016episodic} frame the problem as a \textit{greedy 
MDP} and train a centralised controller to select actions for each agent in a 
sequential fashion. 
\citet{sukhbaatar2016learning} and \citet{peng2017multiagent} train factorised 
but centralised controllers that use special network architectures to share 
information between agents. These approaches assume unlimited bandwidth for 
communication.

One way to decentralise the agents' policies is to learn a separate $Q$-function for each agent as in \emph{Independent $Q$-Learning} \citep{tan1993multi}. \citet{foerster2017stabilising} and \citet{omidshafiei2017deep} examine the problem of instability that arises from the nonstationarity of the environment induced by both the agents' exploration and their changing policies. \citet{rashid2018qmix} and \citet{sunehag2017value} propose learning a centralised value function that factors into per-agent components. 
\citet{guptacooperative} learn separate policies for each agent in an actor-critic framework, where the critic for each agent conditions only on per-agent information. \citet{foerster2017counterfactual} and  \citet{lowe2017multi} propose a single centralised critic with decentralised actors. 
None of these approaches explicitly learns a policy over joint actions and hence are limited in the coordination they can achieve.

\citet{thomas2014psychology} explore the psychology of common knowledge and coordination. \citet{rubinstein1989electronic} shows that any finite number of reasoning steps, short of the infinite number required for common knowledge, can be insufficient for achieving coordination (see Appendix \ref{sec:common_knowledge}). \citet{korkmaz2014collective} examine common knowledge in scenarios where agents use Facebook-like communication. %
\citet{brafman2003coordinate} use a common knowledge protocol to improve coordination in common interest stochastic games but, in contrast to our approach, establish common knowledge about agents' action sets and not about subsets of their observation spaces.

\citet{aumann1974subjectivity} introduce the concept of a \textit{correlated equilibrium}, whereby a shared \textit{correlation device} helps agents coordinate better. \citet{cigler2013decentralized} examine how the agents can reach such an equilibrium when given access to a simple shared \textit{correlation vector} and a communication channel. \citet{boutilier1999sequential} augments the state space with a coordination mechanism, to ensure coordination between agents is possible in a fully observable multi-agent setting. This is in general not possible in the partially observable setting  we consider. 

\citet{amato2014planning} propose MacDec-POMDPs, which use hierarchically optimal policies that allow agents to undertake temporally extended macro-actions. \citet{liu2017learning} investigate how to learn such models in environments where the transition dynamics are not known. \citet{makar2001hierarchical} extend the MAXQ single-agent hierarchical framework \citep{dietterich2000maxq} to the multi-agent domain. They allow certain policies in the hierarchy to be \textit{cooperative}, which entails learning the joint action-value function and allows for faster coordination across agents. \citet{kumar2017federated} use a hierarchical controller that produces subtasks for each agent and chooses which pairs of agents should communicate in order to select their actions.
\citet{oh_few_2008} employ a hierarchical learning algorithm for cooperative control tasks where the outer layer decides whether an agent should coordinate or act independently, and the inner layer then chooses the agent's action accordingly. 
In contrast with our approach, these methods require communication during execution and some of them do not test on sequential tasks.

\citet{nayyar_decentralized_2013} show that common knowledge can be used to reformulate decentralised planning problems as POMDPs to be solved by a central coordinator using dynamic programming. However, they do not propose a method for scaling this to high dimensions. By contrast, MACKRL is entirely model-free and learns trivially decentralisable control policies end-to-end.

\citet{guestrin_context-specific_2002} represent agents' value functions as a sum of context-specific value rules that are part of the agents' fixed a priori common knowledge. By contrast, MACKRL learns such value rules dynamically during training and does not require explicit communication during execution.

Despite using a hierarchical policy structure, MACKRL is not directly related to the family of hierarchical reinforcement learning algorithms \citep{vezhnevets_feudal_2017}, as it does not involve temporal abstraction.

%% file: core/8_conclusion.tex
\section{Conclusion and Future Work}
\label{sec:conclusion}
This paper proposed a way to use common knowledge to improve the ability of decentralised policies to coordinate. To this end, we introduced \method, 
an algorithm which allows a team of agents to learn a fully decentralised policy that nonetheless can select actions jointly 
by using the common knowledge available. 
\method~uses a hierarchy of controllers that can either select joint actions 
for a pair or delegate to independent controllers.

In evaluation on a matrix game and a challenging multi-agent version of 
StarCraft II micromanagement, \method~outperforms strong baselines and even exceeds the state of the art by exploiting common knowledge.
We present approximate versions of \method~that can scale to greater 
numbers of agents and demonstrate robustness to observation noise.

In future work, we would like to further increase \method's scalability and robustness to sensor noise, explore off-policy variants of \method~and investigate how to exploit limited bandwidth communication in the presence of common knowledge. We are also interested in utilising SIM2Real transfer methods \citep{tobin_domain_2017,tremblay_training_2018} in order to apply \method~to autonomous car and unmanned aerial vehicle coordination problems in the real world.

%% file: core/ack.tex
\subsubsection*{Acknowledgements}
We would like to thank Chia-Man Hung, Tim Rudner, Jelena Luketina, and Tabish Rashid for valuable discussions.
This project has received funding from the European Research Council (ERC) under the European Union’s Horizon 2020 research and innovation programme (grant agreement number 637713), the National Institutes of Health (grant agreement number R01GM114311), EPSRC/MURI grant EP/N019474/1 and the JP Morgan Chase Faculty Research Award. This work is linked to and partly funded by the project Free the Drones (FreeD) under the Innovation Fund Denmark and Microsoft. It was also supported by the Oxford-Google DeepMind Graduate Scholarship and a generous equipment grant from NVIDIA.

%% file: core/9_supplementary.tex
\section{Pairwise MACKRL}
This section aims to give a better intuition about Pairwise 
\method~using the example of three agents $\Set A := \{1,2,3\}$.
The explicitly written-out joint policy of Pairwise MACKRL for these agents is:
{\footnotesize
\begin{align*}
\boldsymbol{\pi}_\theta(u^1_\env, u^2_\env, u^3_\env) 
	&=  \pi_{\text{ps},\theta}^{\Set A}(u_\text{ps}^{\Set A}{=}\{\{1\},\{2,3\}\}|\mathcal{I}_s^{1,2,3})  \cdot \pi^1_\theta(u^1_\env |\tau^1) \\[-1mm]
	&\qquad \cdot \Big( 
		\pi^{2,3}_{\text{pc},\theta}(u^{2,3}_\env |\mathcal{I}_s^{2,3})
		+ \pi^{2,3}_{\text{pc},\theta}(u^{2,3}_d |\mathcal{I}_s^{2,3})
			\cdot \pi^2_\theta(u^2_\env |\tau^2) \cdot \pi^3_\theta(u^3_\env |\tau^3)
	\Big)
\\[1mm]
	\quad &+ \pi_{\text{ps},\theta}^{\Set A}(u_\text{ps}^{\Set A}{=}\{\{2\},\{1,3\}\}|\mathcal{I}_s^{1,2,3}) \cdot \pi^2_\theta(u^2_\env |\tau^2 )
	\\[-1mm]
	&\qquad \cdot \Big( 
		\pi^{1,3}_{\text{pc},\theta}(u^{1,3}_\env |\mathcal{I}_s^{1,3})
		+ \pi^{1,3}_{\text{pc},\theta}(u^{1,3}_d |\mathcal{I}_s^{1,3})
			\cdot \pi^1_\theta(u^1_\env |\tau^1) \cdot \pi^2_\theta(u^3_\env |\tau^3)
	\Big) 
\\[1mm]
	\quad &+ \pi_{\text{ps},\theta}^{\Set A}(u_{\text{ps}}^{\Set A}{=}\{\{3\}, \{1,2\}\}|\mathcal{I}_s^{1,2,3} ) \cdot \pi^3_\theta(u^3_\env |\tau^3) \\[-1mm]
	&\qquad \cdot \Big( 
		\pi^{1,2}_{\text{pc},\theta}(u^{1,2}_\env |\mathcal{I}_s^{1,2})
		+ \pi^{1,2}_{\text{pc},\theta}(u^{1,2}_d |\mathcal{I}_s^{1,2})
			\cdot \pi^1_\theta(u^1_\env |\tau^1) \cdot \pi^2_\theta(u^2_\env |\tau^2)
	\Big) 
\,.
\end{align*}
}\!
Conditional variables beyond common knowledge $\mathcal{I}^\mathcal{G}$ 
and action-observation histories $\tau^a$ have been omitted for brevity. 
See Table \ref{tab:pairwise} for a detailed depiction 
of Pairwise \method's hierarchical controllers.

\begin{table}[h]
\hspace{-1.2cm}
\vspace{-3mm}
\begin{center}
{\renewcommand{\arraystretch}{1.3}
 \begin{tabular}{||c l c||} 
 \hline
 Level & Policy / Controller & \#$\pi$ \\
 \hline\hline
 1 & $\pi_{\text{ps}}( u^{\text{ps}} | \mathcal{I}_{s_t}^{\mathcal{A}}, 
 u^{\text{ps}}_{t-1},  h^{\text{ps}}_{t-1})$  & 1 \\
 \hline
 2 & $\pi^{a a'}_{\text{pc}}( u^{aa'} | \mathcal{I}_{s_t}^{a a'}, u^{a a'}_{ 
 t-1},  h^{a a'}_{ t-1 }, aa' ) $ & 3 \\
 \hline
 3 & $\pi^a(u^a | z^a_t, h^{a}_{t-1}, u^{a}_{t-1}, a )$ & 3 \\
\hline
\end{tabular}
}
\end{center}
\caption{Hierarchy of pairwise \method, 
where $h$ is the hidden state of RNNs 
and $z^a_t$ are observations.
\#$\pi$ shows the number of controllers 
at this level for 3 agents.}\label{tab:pairwise}
\end{table} 

However the sampling of each agent's actions $u^a_\env \in \Set U^a$ 
only needs to traverse one branch of the tree, 
as shown in Figure \ref{fig:pairwiseMACKRLsampling}.
At the top level, an agent id partition $u_{\text{ps}}$ is categorically sampled from the pair selector policy $\pi_{\text{ps},\theta}$. At the second level, the pair selector policy for the pair contained in the partition $\pi_{\text{pc}}^{a,b}$ is categorically sampled from in order to receive $u^{a,b}$ where $a,b\in u_{\text{ps}}$. If the delegation action $d$ is sampled, then both $u^a$ and $u^b$ are categorically resampled from their respective independent policies $\pi^a$ and $\pi^b$. Otherwise, $u^a$ and $u^b$ are determined by $u^{a,b}$. The leftover agent $c\not\in u_{\text{ps}}$ samples its action from its corresponding independent policy $\pi^c$. Note that this sampling scheme naturally generalised for $n>3$.

\tikzstyle{level 1}=[level distance=1cm, sibling distance=1cm]
\tikzstyle{level 2}=[level distance=1.5cm, sibling distance=1.5cm]

\tikzstyle{bag} = [text width=4em, text centered]
\tikzstyle{end} = [circle, minimum width=3pt,fill, inner sep=0pt]

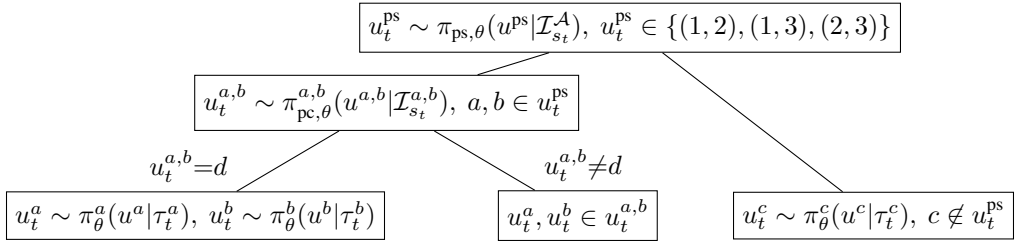
\begin{figure*}[h]
	\begin{center}
\begin{tikzpicture}
    [every tree node/.style={draw},
   level distance=0.3cm,sibling distance=0.5cm, 
   edge from parent path={(\tikzparentnode) -- (\tikzchildnode)}]
\Tree [.\node (foo) {$u^{\text{ps}}_t\sim\pi_{\text{ps},\theta}(u^\text{ps}|\mathcal{I}^\mathcal{A}_{s_t}),\ u^{\text{ps}}_t\in\{(1,2), (1,3), (2,3)\}$}; 
    \edge node[auto=right] {};
    [.\node{$u^{a,b}_t\sim\pi^{a,b}_{\text{pc},\theta}(u^{a,b}|\mathcal{I}^{a,b}_{s_t}),\ a,b \in u^{\text{ps}}_t$};
      \edge node[pos=1.0, auto=right] {$u^{a,b}_t{=}d$};  
      [.\node{$u^a_t\sim\pi^a_\theta(u^a|\tau^a_t),\ u^b_t\sim\pi^b_\theta(u^b|\tau^b_t)$};
      ]
      \edge node[pos=1.0, auto=left] {$u^{a,b}_t{\neq}d$};  
      [.\node{$u^a_t, u^b_t \in u^{a,b}_t$};
      ]
    ]
    \edge[draw=none] node[auto=left] {};  
      [
       \edge[draw=none] node[auto=right] {};  
      [.\node (qux) {$u^c_t\sim\pi^c_\theta(u^c|\tau^c_t),\ c\not\in u^{\text{ps}}_t$};]   ]
    ];
\draw (qux) -- (foo) node[midway,auto=right] {};
\end{tikzpicture}
	\end{center}
	\vspace{-3mm}
	\caption{\label{fig:pairwiseMACKRLsampling} Action sampling for MACKRL for $n=3$ agents.}
	\vspace{-3mm}
\end{figure*}

\newpage

\tikzstyle{level 1}=[level distance=3.5cm, sibling distance=3.5cm]
\tikzstyle{level 2}=[level distance=3.5cm, sibling distance=2cm]

\tikzstyle{bag} = [text width=4em, text centered]
\tikzstyle{end} = [circle, minimum width=3pt,fill, inner sep=0pt]

\begin{figure*}[t!]
	\begin{center}
\begin{tikzpicture}[grow=right, sloped, scale=0.7]
\node[bag] {Game matrix $A,B$}
    child {
        node[bag] {CK available $\text{Yes}, \text{No}$}        
            child {
                node[end, label=right: {$c_1, c_2=\hcancel{\mathcal{CK}}; \sigma_1, \sigma_2=
                \begin{cases} B, B & \text{ with prob. }  p_{\sigma}^2 \\ 
                B, ?  & \text{ with prob. }  \left(1-p_{\sigma}\right)p_{\sigma} \\
                ?, B  & \text{ with prob. }  p_{\sigma}\left(1-p_{\sigma}\right) \\
                ?,?  & \text{ with prob. }  \left(1-p_{\sigma}\right)^2 \\ \end{cases}$}] {}
                edge from parent
                node[above] {$\text{\quad No}$}
                node[below]  {$1 - p_{ck}$}
            }
            child {
                node[end, label=right:
                    {$c_1, c_2=\mathcal{CK}; \sigma_1,\sigma_2=B$}] {}
                edge from parent
                node[above] {$\text{Yes}$}
                node[below]  {$p_{ck}$}
            }
            edge from parent 
            node[above] {$B$}
            node[below]  {$\frac{1}{2}$}
    }
    child {
        node[bag] {CK available $\text{Yes}, \text{\quad No}$}        
        child {
                node[end, label=right:
                    {$c_1, c_2=\hcancel{\mathcal{CK}}; \sigma_1, \sigma_2=
                \begin{cases} A, A & \text{ with prob. }  p_{\sigma}^2 \\ 
                A, ?  & \text{ with prob. }  \left(1-p_{\sigma}\right)p_{\sigma} \\
                ?, A  & \text{ with prob. }  p_{\sigma}\left(1-p_{\sigma}\right) \\
                ?, ?  & \text{ with prob. }  \left(1-p_{\sigma}\right)^2 \\ \end{cases}$}] {}
                edge from parent
                node[above] {$\text{\quad No}$}
                node[below]  {$1-p_{ck}$}
            }
            child {
                node[end, label=right:
                    {$c_1, c_2=\mathcal{CK}; \sigma_1,\sigma_2=A$}] {}
                edge from parent
                node[above] {$\text{Yes}$}
                node[below]  {$p_{ck}$}
            }
        edge from parent         
            node[above] {$A$}
            node[below]  {$\frac{1}{2}$}
    };
\end{tikzpicture}
	\end{center}
	\vspace{-3mm}
	\caption{\label{fig:delegation} Probability tree for our simple single-step matrix game. The game chooses randomly between matrix $A$ or $B$, and whether common knowledge is available or not. If common knowledge is available, both agents can condition their actions on the game matrix chosen. Otherwise, both agents independently only have a random chance of observing the game matrix choice. Here, $p_{ck}$ is the probability that common knowledge exists and $p_\sigma$ is the probability that an agent independently observes the game matrix choice. The observations of each agent $1$ and $2$ are given by tuples $(c_1, \sigma_1)$ and $(c_2, \sigma_2)$, respectively, where $c_1, c_2\in\{\mathcal{CK},\hcancel{\mathcal{CK}}\}$ and $\sigma_1, \sigma_2\in\{A,B,?\}$. }
	\vspace{-3mm}
\end{figure*}
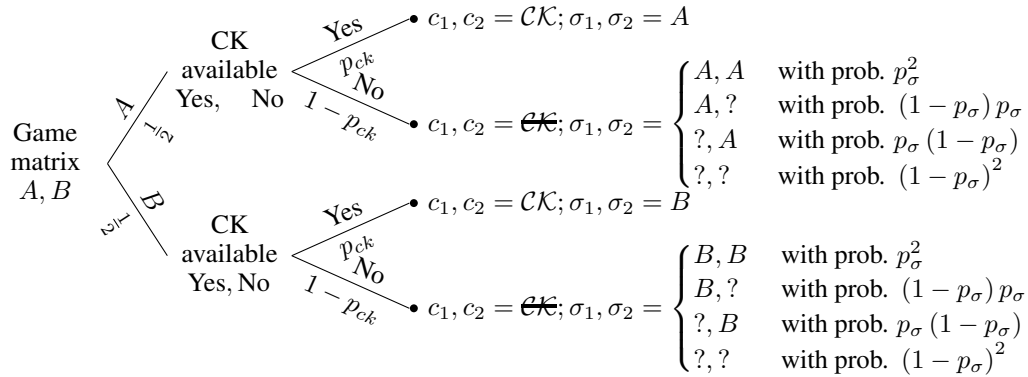

\section{Experimental Setup - StarCraft II}
\label{sec:sc2_exp_supp}

All policies are implemented as two-layer recurrent neural networks (GRUs) with 
$64$ hidden units, while the critic is feed forward and uses full state 
information. Parameters are shared across controllers within each of the second 
and third levels of the hierarchy. We also feed into the policy the agent index 
or index pairs. For exploration, we use a bounded softmax 
distribution in which the agent samples from a softmax over the policy logits 
with probability $(1-\epsilon) $ and samples randomly with probability 
$\epsilon$. We anneal $\epsilon$ from $0.5$ to $0.01$ across 
the first $50$k environment steps.

Episodes are collected using eight parallel SCII environments. Optimisation is 
carried out on a single GPU with Adam and a learning rate of $0.0005$ for both 
the agents and the critic.
The policies are fully unrolled and updated in a large 
mini-batch of $T \times B$ entries, where $T=60$ and $B = 8$. 
By contrast, the critic is optimised in small mini-batches of size $8$, one for 
each time-step, looping backwards in time. We found that this stabilised 
and accelerated training compared to full batch updates for the critic. The 
target network for the critic is updated after every $200$ critic updates.
We use $\lambda=0.8$ in TD($\lambda$) to accelerate reward propagation.

\section{Pair controller introspection}
\label{sec:pair_intro}

\begin{wrapfigure}{R}{0.6\textwidth}
	\begin{center}
		\includegraphics[width=0.6\textwidth]{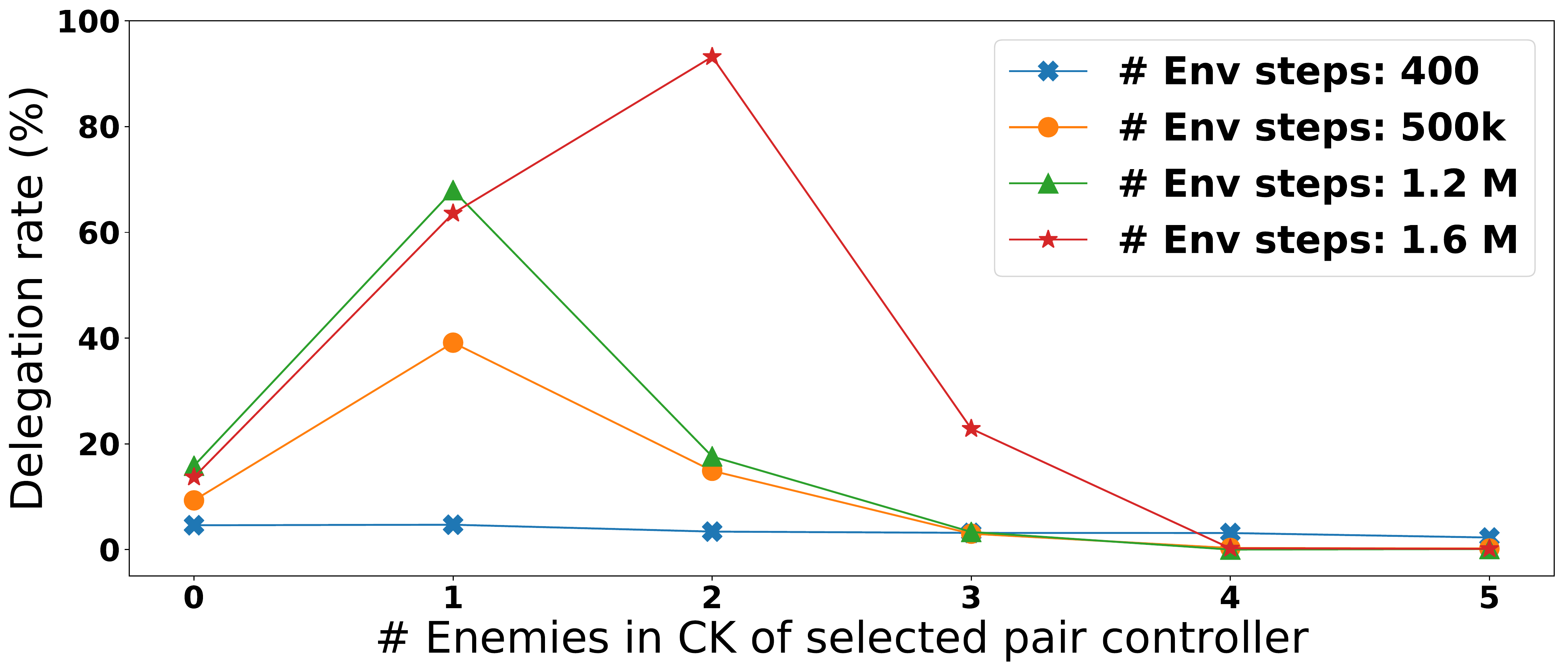}
	\end{center}
	\vspace{-3mm}
	\caption{\label{fig:delegation} Delegation rate 
	vs. number of enemies (2s3z) 
	in the common knowledge of the pair controller over training. }
	\vspace{-3mm}
\end{wrapfigure}

We now test the hypothesis that \method's superior performance is indeed due to its ability to learn how to use common knowledge for coordination.
To demonstrate that the pair controller can indeed learn to delegate strategically,
we plot in Figure \ref{fig:delegation} 
the percentage of delegation actions $u_d$
against the number of enemies in the common knowledge of the selected pair controller, in situations where there is at least some common knowledge.

Since we start with randomly initialised policies, at the beginning of training 
the pair controller delegates only rarely to the decentralised controllers. 
As training proceeds, it learns to delegate in most situations where the number of enemies in the common knowledge of the pair is small, the exception being no visible enemies, which happens too rarely ($5\%$ of cases). 
This shows that MACKRL can learn to delegate in order to take advantage of the private observations of the agents, but also learns to coordinate in the joint action space when there is substantial common knowledge.

\section{Holenstein's Strategy}
\label{sec:holenstein}

Given two agent-specific pair controller policies $\tilde{\pi}^{aa'}_{pc,a}$, both agents can optimally minimise disagreement when sampling independently form their respective policies by following Holenstein's strategy \citep{holenstein_parallel_2007,bavarian_optimality_2016}:
With a suitably chosen $\gamma>0$, each agent $a$ is assigned a set 

\begin{equation} \mathcal{H}_a=\{(u, p)\in\mathcal{U}^{aa'}_{pc}\times\Gamma:p<\tilde{\pi}^{aa'}_{pc,a}(u)\},\ \Gamma=\{0,\gamma,2\gamma,\dots,1\}
\end{equation}

Let $\zeta$ be a shared $\xi$-seeded random permutation of the elements in $\mathcal{U}^{aa'}_{pc}\times\Gamma$, then agent $a$ samples $\zeta(i_a)$, where $i_a$ is the smallest index such that $\zeta(i_a)\in\mathcal{H}_a$ (and agent $a'$ proceeds analogously). Given the total variational distance $\delta$ between the categorical probability distributions defined by $\tilde{\pi}^{aa'}_{pc,a}$ and $\tilde{\pi}^{aa'}_{pc,a'}$, the disagreement probability of agents $a,a'$ is then guaranteed to be at most $2\delta/(1+\delta)$ \citep{bavarian_optimality_2016}. 

%% file: core/common_knowledge.tex
\section{Common Knowledge with Entities}
\label{sec:common_knowledge}

To exploit a particular form of 
{\em field-of-view common knowledge} with MACKRL, 
we formalise an instance of a Dec-POMDP, 
in which such common knowledge naturally arises.
In this Dec-POMDP, the state $s$ is composed of a number of entities $e \in 
\Set E$, with state features $s^e$, i.e., $s = \{s^e \,|\, e \in \Set E\}$. 
Some entities are agents $a \in \Set A \subseteq \Set E$.  
Other entities could be %
enemies, obstacles, or goals.

The agents have a particular form of partial observability: the observation 
$z^a$ contains the subset of state features $s^e$ from all the entities $e$ 
that $a$ can see. Whether $a$ can observe $e$ is determined
by the binary mask $\mu^a\big(s^a, s^e\big) \in \{\top, \bot\}$ over the 
agent's and entity's observable features. An agent can always observe itself,
i.e., $\mu^a(s^a, s^a) = \top, \forall a \in \Set A$.
The set of all entities the agent can see is therefore
$\Set M^a := \{ e \,|\, \mu^a(s^a, s^e) \} \subseteq \Set E$,
and the agent's observation is specified by the deterministic observation 
function $o(s,a)$ such that
$z^a = o(s,a) = \{s^e \,|\, e \in \Set M^a\} \in \Set Z$.
In the example of Figure \ref{fig:common_info},
$\Set M^A = \Set M^B = \{A,B\}$ and $\Set M^C = \{A,B,C\}$.

This special Dec-POMDP yields perceptual aliasing in which the state features 
of each entity are either accurately observed or completely occluded.
The Dec-POMDP could be augmented with additional state features that do not 
correspond to entities, as well as additional possibly noisy observation 
features, without disrupting the common knowledge we establish about entities.  
For simplicity, we omit such additions.

A key property of the binary mask $\mu^a$ is that it depends only 
on the features $s^a$ and $s^e$ to determine 
whether agent $a$ can see entity $e$. 
If we assume that an agent $a$'s mask $\mu^a$ is common knowledge, 
then this means that another agent $b$, that can see $a$ and $e$, 
i.e., $a,e \in \Set M^b$, can deduce whether $a$ can also see $e$.  
This assumption can give rise to common knowledge about entities.
Figure~\ref{fig:common_info} demonstrates this for
3 agents with commonly known observation radii.

The \emph{mutual knowledge} $\Set M^\g$ of a group of agents 
$\g \subseteq \Set A$ in state $s$ is the set of entities that all agents in 
the group can see in that state:
$\Set M^\g := \setintersect_{a \in \g} \Set M^a$.  However, mutual knowledge 
does not imply common knowledge.
Instead, the {\em common knowledge} $\Set I^\g$ 
of group $\g$ in state $s \in \Set S$ 
is the set of entities such that
 all agents in $\g$ see $\Set I^\g$, 
 know that all other agents in $\g$ see $\Set I^\g$, 
 know that they know that all other agents see $\Set I^\g$,
and so forth \citep{osborne1994course}.

To know that another agent $b$ also sees $e \in \Set E$,
agent $a$ must see $b$ and $b$ must see $e$,
i.e., $\mu^a(s^a, s^{b}) \wedge \mu^{b}(s^{b}, s^e)$.  %
Common knowledge $\Set I^\g$ can then be formalised recursively
for every agent $a \in \g$ as:
\vspace{-1mm}
\begin{equation} \label{eq:common_knowledge_definition}
	\Set I^{a}_0 \;:=\; \Set M^a \,, 
	\qquad\quad
	\Set I^{a}_m \;:=\; \Setintersect_{b \in \Set G} 
		\big\{ e \in \Set I^{b}_{m-1} \,|\, \mu^a(s^a, s^b) \big\} \,, 
	\qquad\quad
	\Set I^\g \;:=\; \lim_{m\to \infty} \Set I^{a}_m \,.
\end{equation}

\vspace{-3mm}\noindent
This definition formalises the above description 
that common knowledge is the set of entities 
that a group member sees ($m=0$), 
that it knows all other group members see as well ($m=1$), 
and so forth ad infinitum. 
In the example of Figure \ref{fig:common_info},
$\Set I^{AB} = \{A,B\}$ and 
$\Set I^{AC} = \Set I^{BC} = \Set I^{ABC} = \varnothing$.

The following lemma establishes that, in our setting, if a group of agents can 
all see each other, their common knowledge is their mutual knowledge.

\begin{lemma}
In the setting described in this Section, 
and when all masks are known to all agents,
the common knowledge of a group of agents $\Set G$ in state $s \in \Set S$ is 
\begin{equation} \label{eq:common_knowledge_agent} 
	\Set I^{\Set G} 
	\;\;=\;\; \left\{ \begin{array}{cl}
		\Set M^\g, & \text{if} 
		\Setand\limits_{a, b \, \in \Set G} \mu^a(s^a, s^{b})  
	\\
		\varnothing, & \text{otherwise} 
	\end{array} \right.\,. 
\end{equation}
\label{eq:lemma}
\end{lemma} 
\vspace{-6mm}
\begin{proof}
	The lemma follows by induction on $m$.
	The recursive definition of common knowledge
	(\ref{eq:common_knowledge_definition})
	holds trivially if $\Set I^\g = \varnothing$.
	Starting from the knowledge of any agent $a$ in state $s$, 
	$\Set I_0^{a} = \Set M^a$,
	definition (\ref{eq:common_knowledge_definition}) yields:
	$$
		\Set I^{a}_1 = 	\left\{ \begin{array}{cl}
			\Set M^\g, & \text{if} \Setand\limits_{b \in \g}\mu^a(s^a,s^b) \\
			\varnothing, & \text{otherwise} 
		\end{array} \right. \,.
	$$
	Next we show inductively that if all agents in group $\g$
	know the mutual knowledge $\Set M^\g$ of state $s$
	at some iteration $m$, that is, 
	$\Set I^{c}_m \stackrel{\scriptstyle \text{ind.}}{=} \Set M^\g$,
	then this mutual knowledge 
	becomes common knowledge two iterations later.
	Applying the definition (\ref{eq:common_knowledge_definition})
	for any agent $a \in \g$ twice yields:
	\begin{eqnarray*}
		\Set I^{a}_{m+2} \kern-1.5ex&=&\kern-1.5ex
			\Setintersect\limits_{b \in \g} \Setintersect\limits_{c \in \g}
			\Big\{e \in \Set I^{c}_m \Big| 
				\mu^a(s^a,s^b) \wedge \mu^b(s^b,s^c) \Big\} %
		=
			\scriptstyle \big\{ e \in \Set E  \,\big|\, 
			\Setand\limits_{b \in \g}\big(\mu^a(s^a, s^b) \,
			\wedge \Setand\limits_{c \in \g} \big(\mu^b(s^b, s^c) 
			\,\wedge\, e \in \Set I^{c}_m \big) \big)\big\} \\
		&\stackrel{\text{ind.}}{=}&\kern-1.5ex 
			\textstyle \big\{ e \in \Set M^\g  \,\big|\, 
			\Setand\limits_{b,c \in \g} \mu^b(s^b, s^c) \big\} \,,
	\end{eqnarray*} 
	
	\vspace{-4mm}\noindent
	which is the right side of (\ref{eq:common_knowledge_agent}),
	and where we used
	$$ \textstyle
		\Setand\limits_{b \in \g}\kern-1ex\big(\mu^a(s^a, s^b) \,
		\wedge \Setand\limits_{c \in \g}\kern-0.5ex\mu^b(s^b, s^c)  \big)
		= \kern-1ex\Setand\limits_{b,c \in \g}\kern-0.5ex \mu^b(s^b, s^c)  \,, 
		\; \forall a \in \Set G\,.
	$$
	Finally, applying (\ref{eq:common_knowledge_definition}) 
	one more time to this result, yields:
	\begin{eqnarray*}
		\Set I^a_{m+3} &=& \Setintersect\limits_{b \in \g}
			\Big\{ e \in \Set I^b_{m+2} \Big| \mu^a(s^a, s^b) \Big\}
		\quad=\quad \Set I^a_{m+2} \,.
	\end{eqnarray*}
	For all $m \geq 3$, $\Set I^a_m$ remains thus 
	the right hand side of (\ref{eq:common_knowledge_agent}).
	As $\Set I^\g = \lim_{m\to \infty} \Set I^a_m$,
	we can thus conclude that,
	starting at the knowledge of any agent of group $\g$,
	in which all agents can see each other,
	the mutual knowledge is	the common knowledge.
\end{proof}

The common knowledge can be computed 
using only the visible set $\Set M^a$ of every agent $a \in \Set G$.
Moreover, actions that have been chosen 
by a policy, which itself is common knowledge, and that further depends only on 
common knowledge 
and a shared random seed, are also common knowledge.
The common knowledge of group $\g$ up to time $t$ is thus 
some common prior knowledge $\xi$ 
and the commonly known trajectory 
$\tau_t^{\Set G} = (\xi, z^{\Set G}_1, \vu^\g_1, 
\ldots, z^{\Set G}_t, \vu^\g_t)$, 
with $z_k^{\Set G} = \{s_k^e \,|\, e \in \Set I^{\Set G}\}$.
Knowing all binary masks $\mu^a$ makes it possible to derive $\tau_t^{\g} = 
\Set I^\g(\tau^a_t, \xi)$ from the observation
trajectory $\tau_t^a = (z^a_1, \ldots, z^a_t)$ of 
any agent $a \in \Set G$ and the shared prior knowledge $\xi$.
A function that conditions on $\tau^{\Set G}$ 
can therefore be computed independently by every member of $\Set G$.

Note that (by definition)
common knowledge can only arise from entities 
that are observed {\em identically} by all agents.
If only one agent receives non-deterministic observations, 
for example induced by sensor noise,
the other agents cannot deduce the group's mutual (and thus common) knowledge.
Our method therefore only guarantees perfect decentralisation of the learned 
policy in settings 
with deterministic observations, like simulations and computer games.
However, in Section \ref{sec:matrix_results} we show empirically that, 
using a naive correlated sampling protocol similar 
to the theoretically optimal Holenstein protocol 
\citep{holenstein_parallel_2007,bavarian_optimality_2016}, 
\method~can still succeed in the presence of moderate sensor noise.